\documentclass{paper}
\usepackage{amsfonts,amsmath,latexsym,amscd,graphicx,amssymb}

\newcommand{\be}{\begin{equation}}
\newcommand{\ee}{\end{equation}}
\newcommand{\ba}{\begin{eqnarray}}
\newcommand{\ea}{\end{eqnarray}}

\newcommand{\pa}{\partial}
\newcommand{\f}{\frac}

\begin{document}

\title{Small-amplitude capillary-gravity water
waves:  exact solutions and  particle motion  beneath such waves}

\author{\normalsize Delia IONESCU-KRUSE\\
\normalsize Institute of Mathematics of
the Romanian Academy,\\
\normalsize P.O. Box 1-764, RO-014700, Bucharest,
 Romania\\
\normalsize E-mail: Delia.Ionescu@imar.ro\\[10pt]}

 \date{}

\maketitle

\begin{abstract}
Two-dimensional periodic surface waves propagating under the
combined influence of gravity and surface tension  on water of
finite depth are considered. Within the framework of
small-amplitude waves, we find the exact solutions of the
nonlinear differential equation system which describes the
particle motion in the considered case, and we describe the
possible particle trajectories.  The required computations involve
elliptic integrals of the first kind, the Legendre normal form and
a solvable Abel differential equation of the second kind. Some
graphs of the results are included.
\end{abstract}

\section{Introduction}

We are concerned with the motion of periodic plane waves which are
propagated in water of finite depth and with free surface, under a
gravitational field.  The governing equations of motion  are the
incompressible Euler equations. On the free surface we take into
account the influence of surface tension. The surface tension will
appear in the formulation of the boundary conditions. Thus, one
obtains the boundary-value problem for capillary-gravity waves.
There are very few explicit solutions known for the water-wave
problems. The first such solution for pure gravity water waves,
was described by Gerstner \cite{gerstner}. This solution was
independently rediscovered later by Rankine \cite{rankina}. Modern
detailed descriptions of this wave are given in  recent papers
\cite{c2001a} and \cite{henry4}. Gerstner's solution is
restrictive: it exists only for deep water and it describes a
rotational wave. Related to Gerstner's solution, Constantin, in
\cite{c2001b}, constructed   an explicit rotational solution to
the nonlinear governing equations describing gravity water waves
that progress along the shoreline. Beneath Gerstner's waves it is
possible to have a motion of the fluid where all particles
describe circles with a depth-dependent radius (\cite{c2001a},
\cite{henry4}). Crapper \cite{crapper} derived an exact solution
for pure capillary waves travelling at a constant velocity at the
surface of a fluid of infinite depth. He showed that as the
amplitude increases, the waves develop broad crests and sharp
troughs. Crapper's solutions ultimately reach a limiting form for
which there is a trapped bubble at the troughs. For higher values
of the amplitude the free-surface profiles are self-intersecting
and therefore non physical. In \cite{hogen1}, by the use of the
general method for calculating trajectories given by
Longuet-Higgins \cite{longuet}, the particle trajectories in
Crapper's nonlinear capillary waves  are derived. It is found that
the orbits of the steeper waves are neither circular nor closed.
Crapper's solution was extended in the case of a fluid of finite
depth by Kinnersley \cite{kinn}. The particle trajectories for
Kinnersley's waves were calculated in \cite{hogen3}. For nonlinear
capillary-gravity waves no exact analytic solution has yet been
found. Making use of numerical studies, in \cite{hogen2} the
particle trajectories in irrotational nonlinear capillary-gravity
waves  are investigated. There are several unanswered questions
about water waves with surface tension and vorticity. The
existence of regular periodic travelling waves with vorticity was
recently established (see \cite{cs}, \cite{w}). For steady
periodic gravity waves the symmetry is known to be ubiquitous (see
\cite{cew}, \cite{hur}). The study of the symmetry of rotational
water waves was initiated in \cite{CE1}, \cite{CE2}; for
irrotational flows see also \cite{okamoto}. However, exact
information about the flow beneath such waves is not readily
available, even in the irrotational case. This paper addresses
this issue.

It is fortunate that a great number of observations can be
explained on the basis of the small-amplitude wave theory. Within
this framework,  in this paper we investigate the
capillary-gravity waves and the
 internal motion of the fluid under the passage of such waves.
 We simplify the full system
of equations by a linearization
 found in \cite{io},  which is around still water and
which  is slightly different from the classical case in line with
the Stokes condition for irrotational flows (for the latter, see
for example, \cite{cev} and \cite{cv} where perturbations of
laminar flows are considered, that is, flows characterized by a
flat surface $\eta=0$; these two approaches to linearization are
surveyed in \cite{ev2}). After rewriting the governing equations
of motion and the boundary conditions for the capillary-gravity
water-wave problem in an appropriate non-dimensional form, three
non-dimensional parameters arise: $\epsilon=$amplitude/depth, the
amplitude parameter, $\delta=$depth/wavelenght, the shallowness
parameter and  $W_e$, called Weber number, which comes from the
surface tension on the free surface. These parameters help us to
characterize the various types of approximation. We  suppose that
the water flow is irrotational; thus, in addition to the full
system of equations we
 also have the irrotational condition that we write in the suitable set of
non-dimensional variables. The linearized problem is obtained by
letting $\epsilon\rightarrow 0$, $\delta$ and $W_e$ being fixed.
Solving this problem,  we obtain a parameter $c_0$ by which we can
describe different backward flows in the irrotational case:
 still water ($c_0=0$), favorable uniform current $c_0>0$ and  adverse
uniform current $c_0<0$. After getting the general solution of the
linearized water-wave problem  we investigate further the
nonlinear equations of the motion of the fluid particles  to offer
much insight into the fluid motion.

It was widely believed that as the small-amplitude waves propagate
on the surface the particles of the fluid move on  closed orbits.
Analyzing the first-order approximation of the nonlinear ordinary
differential equation system which describes the particle motion,
it was indeed obtained  that all water particles trace closed,
circular or elliptic, orbits (see, for example, \cite{debnath},
\cite{johnson-carte}, \cite{lamb}, \cite{lighthill},
\cite{sommerfeld}, \cite{stoker}). But in \cite{cv} it was proved,
using phase-plane considerations for the nonlinear ordinary
differential equation system which describes the particle motion,
that in linear irrotational gravity water waves no particle
trajectory is actually closed, unless the free surface is flat.
Each particle trajectory involves over a period a backward/forward
movement, and the path is an elliptical arc with a forward drift;
on the flat bed the particle path degenerates to a
backward/forward motion. In \cite{ev2}  recent results in the
understanding of particle paths within different types of
progressive water waves are  surveyed, in the framework of linear
theory as well as in the framework of  exact theory of periodic
symmetric waves, and in the presence or not of the background
currents and vorticity. The results are in agreement with Stokes'
observation for a net mass drift \cite{stokes}. In the linear
framework, by using phase-plane considerations, one obtains that
the particle trajectories in linear deep-water waves (see
\cite{cev}), in linear gravity water waves over finite depth
 with constant vorticity (see \cite{ehrnst}, \cite{ev}), and in
linear irrotational capillary and capillary-gravity water waves
(see \cite{henry2}, \cite{henry3}) are not closed. Small-amplitude
shallow-water waves were studied in \cite{io} (the irrotational
case) and \cite{io2} (the constant vorticity case), and the exact
solutions of the nonlinear ordinary differential equation system
which describes the particle motion below such waves were found.
Depending on the strength of the underlying uniform current
\cite{io} or depending on the relation between the initial data
and the constant vorticity \cite{io2}, it was obtained that some
particle trajectories are undulating curves to the right, or to
the left, others are loops with forward drift, or with backward
drift, and others can follow some peculiar shapes. In the
framework of exact theory, the forward drift of the particles as
the wave progresses was given by analyzing a free-boundary problem
for harmonic functions in a planar domain (see \cite{c2007} for
Stokes waves, \cite{CE3} for solitary waves and \cite{henry} for
deep-water Stokes waves) or by applying local bifurcation theory
(see \cite{w2} for small-amplitude waves with vorticity).

 In this paper  we
continue the study started in \cite{io3}. We provide explicit
solutions for the ordinary differential equation system describing
the motion of the particles beneath small-amplitude
capillary-gravity waves which propagate on the surface of an
irrotational water flow with a flat bottom. In the case when the
constant $c_0$ equals the non-dimensional speed of propagation of
the linear wave, the required computations involve elliptic
integrals of the first kind and their Legendre's normal form. The
six exact solutions obtained in this case contain  Jacobian
elliptic functions in their expressions.  In the case when the
constant $c_0$ is different from the non-dimensional speed of
propagation of the linear wave, the computations involve a
solvable Abel differential equation of the second kind.   In both
cases we remark that the solutions obtained  are not closed
curves. We analyze some solutions  in detail and we draw their
graphs.

\section{Preliminaries}

\subsection{The water-wave problem}

For two-dimensional periodic waves the motion is identical in any
direction parallel to the crest line and is periodic in the
spatial direction in which the wave is propagating. To describe
these waves we consider a cross section of the flow that is
perpendicular to the crest line with Cartesian coordinates
$(x,z)$, the $x$-axis being in the direction of wave propagation
and the $z$-axis pointing vertically upwards.  The water flow
under consideration is bounded by a rigid horizontal surface below
at $z=0$ and a free surface above
 at $z=h_0+\eta(x,t)$, where the constant $h_0>0$ is the mean water level.
 We denote by  $(u(x,z,t), v(x,z,t))$  the velocity of the water.
 Let gravity act now in concert with surface tension.
   The surface tension will play a role in
the formulation of the boundary conditions but not in the
equations of motion valid in the fluid domain. Assuming that the
water is both  homogeneous (constant density $\rho$) (see
\cite{lighthill}) and inviscid, we obtain within the fluid domain
the  equation of mass conservation (MC) together with Euler's
equations (EEs)(see \cite{johnson-carte}).  The boundary
conditions for the water-wave problem are the kinematic boundary
conditions as well as the dynamic boundary condition. The
kinematic boundary conditions (KBCs) express the fact that the
same particles always form the free-water surface and that the
fluid is assumed to be bounded below by a hard horizontal bed
$z=0$. The dynamic boundary condition (DBC) expresses the fact
that the difference of pressure on the two sides of the surface
$\eta$ is balanced by the effects of surface tension.
 Thus, the boundary-value problem
for capillary-gravity water waves is
\begin{equation}
\begin{array}{c}
\begin{array}{c}
u_t+uu_x+vu_z=-\f1{\rho} p_x\\  v_t+uv_x+vv_z=-\f1{\rho} p_z-g\\
\end{array}
\quad \quad \quad \quad \textrm{ (EE) }\\
 \qquad \qquad u_x+v_z=0  \qquad \qquad \qquad \qquad \textrm{ (MC)  }\\
\begin{array}{c}
  v=\eta_t+u\eta_x \, \, \textrm{ on }\,
z=h_0+\eta(x,t)\\
 v=0 \, \,
\textrm { on } z=0
\end{array}
\quad \,\,\, \textrm{ (KBC) }
\\
\qquad
 p=p_0-\f{\Gamma}{R}, \,  \textrm{ on } z=h_0+\eta(x,t)
  \quad \quad \textrm{ (DBC)} \end{array} \label{e+bc}
\ee where  $p(x,z,t)$ is the pressure, $g$ is the constant
gravitational acceleration, $p_0$ is the constant atmospheric
pressure, the parameter $\Gamma(>0)$ is the coefficient of surface
tension and $\f{1}{R}$ is the mean curvature (up to a factor 1/2)
of the surface.
 For a surface defined as the
function $\eta(x,t)$, the mean curvature has the following
expression \be \f{1}{R}=\f{\eta_{xx}}{(1+\eta^2_x)^{3/2}} \ee
 In respect of the well-posedness for the
initial-value problem for (\ref{e+bc}) there has been significant
recent progress; see \cite{shkoller} and the references therein.

An important category of flows is those of zero vorticity
(irrotational flows), characterized by the additional equation \be
u_z-v_x=0 \label{vorticity}\ee In what follows we will consider
this type of flow. The idealization of irrotational flow is
physically relevant in the absence of non-uniform currents in the
water.

\subsection{Small-amplitude approximation of the
  water-wave problem}

We search for a linear approximation of the water-wave problem
(\ref{e+bc})-(\ref{vorticity}). We set the constant water density
$\rho=1$. If $\lambda>0$ is the wavelength and $a>0$ is the wave
amplitude, we make the following change of variables (yielding the
non-dimensionalization of the problem; see \cite{johnson-carte})
\begin{equation}
\begin{array}{c}
x\mapsto\lambda x,  \quad z\mapsto h_0 z, \quad \eta\mapsto a\eta,
\quad t\mapsto\f\lambda{\sqrt{gh_0}}t,\\
  u\mapsto  \sqrt{gh_0}u,
\quad v\mapsto h_0\f{\sqrt{gh_0}}{\lambda}v
\end{array} \label{nondim}\end{equation}
\begin{equation} p\mapsto p_0+ g h_0(1-z)+ g h_0 p
\label{p}\end{equation} where, to avoid new notation, we have used
the same symbols for the non-dimensional variables  $x$, $z$,
$\eta$, $t$, $u$, $v$, $p$ on the right-hand side.
 The non-dimensional pressure
variable  measures the deviation from the hydrostatic pressure
$p_0+ g h_0(1-z)$. We now apply the following scaling of the
non-dimensional variables
\begin{equation} p\mapsto \epsilon p,\quad
(u,v)\mapsto\epsilon(u,v) \label{scaling}\end{equation} avoiding
again the introduction of  new notation. \\
Taking into account (\ref{nondim}), (\ref{p}) and (\ref{scaling}),
the two-dimensional capillary-gravity waves on irrotational water
of finite depth are described, in non-dimensional scaled
variables, by the following boundary-value problem:
\begin{equation}
\begin{array}{cc}
u_t+\epsilon(uu_x+vu_z)=- p_x&\\  \delta^2[v_t+\epsilon(uv_x+vv_z)]=- p_z&\\
 u_x+v_z=0&\\
  u_z-\delta^2v_x=0&\\
  v=\eta_t+\epsilon u\eta_x  \, & \textrm{ on }\,
z=1+\epsilon\eta(x,t)\\
p=\eta-\left(\f{\Gamma}{g\lambda^2}\right)
\f{\eta_{xx}}{(1+\epsilon^2\delta^2\eta^2_x)^{3/2}} \, & \textrm{
on }\,
z=1+\epsilon\eta(x,t)\\
 v=0 \, &
\textrm { on } z=0
 \end{array}
\label{e+bc1''} \end{equation}
  where we have introduced the
amplitude parameter $\epsilon=\f a{h_0}$ and the shallowness
parameter $\delta=\f {h_0}{\lambda}$. It is conventional to write
$\f{\Gamma}{\rho g \lambda^2}=\delta^2W_e$, with
$W_e=\f{\Gamma}{\rho g h_0^2}\,\,$ a Weber number.
This  parameter is used to measure
the size of the surface tension contribution.

By letting $\epsilon\rightarrow 0$, $\delta$ and $W_e$ being
fixed, we obtain a linear approximation of
 our problem, that is,
\begin{equation}
\begin{array}{cc}
u_t+p_x=0&\\ \delta^2v_t+ p_z=0&\\
 u_x+v_z=0&\\
 u_z-\delta^2v_x=0&\\
v=\eta_t  \, & \textrm{ on }\,
z=1\\
  p=\eta-\delta^2 W_e\eta_{xx} \, & \textrm{ on }\,
z=1\\
 v=0 \, &
\textrm { on } z=0
\end{array}
\label{small} \end{equation}
  Manipulating the first four
equations of  system (\ref{small}), we obtain that \be
v_{zz}+\delta^2v_{xx}=0\label{17'}\ee Applying the method of
separation of variables, we seek the  solution of the  equation
(\ref{17'}) in the form \be v(x,z,t)=F(x,t)G(z,t) \label{19}\ee
Substituting (\ref{19}) into the equation (\ref{17'}), separating
the variables and taking into account the expressions of $v$ on
the boundaries, that is, the fifth equation and the last equation
in (\ref{small}), we find \be v(x,z,t) =\f{1}{\sinh(k\delta)}\sinh
(k\delta z)\eta_t\label{22}\ee where $k\geq 0 $ is a constant that
might depend on time.
 For the component $u$ of the
velocity field,
 taking into account (\ref{22}) and the
fourth equation of  system (\ref{small}), we obtain \be
u(x,z,t)=\f{\delta}{k\sinh (k\delta)}\cosh (k\delta
z)\eta_{tx}+\mathcal{F}(x,t) \label{23}\ee where
$\mathcal{F}(x,t)$ is an arbitrary function. The components $u$
and $v$ of the velocity have to fulfill also the third equation in
(\ref{small}), hence, in view of (\ref{22}) and (\ref{23}),  \be
\f{\delta}{k\sinh (k\delta)}\cosh (k\delta
z)\eta_{txx}+\f{\pa\mathcal{F}(x,t)}{\pa x}=-\f{k\delta }{\sinh
(k\delta)}\cosh (k\delta z)\eta_{t} \label{23'}\ee The above
relation must hold for all values of $x\in\mathbf{R}$, and $0\leq
z\leq 1$. It follows that the function $\mathcal{F}(x,t)$ is
independent of $x$, therefore we will denote this function by
$\mathcal{F}(t)$, and \be \eta_{txx}+k^2\eta_t=0 \label{25}\ee We
seek periodic travelling wave solutions; thus, for the equation
(\ref{25}) with \be k=2\pi\ee
  we choose the following solution \be
\eta(x,t)=\cos(2\pi(x-ct)) \label{26}\ee where $c$ represents the
non-dimensional speed of propagation of the linear wave and is to
be determined.\\
 In order to find the expressions of the
pressure we take into account the first two equations in
(\ref{small}) and the expressions of the velocity field from
above. Thus, we obtain  \be p(x,z,t)=\f{2\pi\delta
c^2}{\sinh(2\pi\delta)}\cosh(2\pi\delta z)\cos(2\pi(x-ct))+
x\mathcal{F}'(t) \label{29}\ee  On the free surface $z=1$ the
pressure (\ref{29}) has to fulfill the sixth equation of  system
(\ref{small}). Hence, in view of (\ref{26}), we get \be 2\pi\delta
c^2\coth(2\pi\delta)\cos(2\pi(x-ct))+
x\mathcal{F}'(t)=(1+4\pi^2\delta^2W_e)\cos(2\pi(x-ct)) \ee The
above relation must hold for all values $x\in \mathbf{R}$;
therefore, we get \be \mathcal{F}(t)=\textrm{constant}:=c_0\ee and
we provide the non-dimensional speed of the linear wave \be
c^2=\f{\tanh(2\pi\delta)}{2\pi\delta}(1+4\pi^2\delta^2W_e)=
\f{\lambda}{2\pi h_0}\left(1+
\f{4\pi^2\Gamma}{g\lambda^2}\right)\tanh\left(\f{2\pi
h_0}{\lambda}\right)\label{c}\ee We observe thus, that the speed
of propagation of the wave varies with the wavelength $\lambda$,
with the undisturbed depth $h_0$ and with the coefficient of
surface tension $\Gamma$.

Summing up,   system (\ref{small}) has the solution \be
 \begin{array}{llll}
 \eta(x,t)=\cos(2\pi(x-ct))\\
u(x,z,t)=\f{2\pi\delta c}{\sinh (2\pi\delta)}\cosh (2\pi\delta
z)\cos(2\pi(x-ct))+c_0\\
v(x,z,t)=\f{2\pi c}{\sinh(2\pi\delta)}\sinh (2\pi\delta
z)\sin(2\pi(x-ct))\\
 p(x,z,t)=\f{2\pi\delta
c^2}{\sinh(2\pi\delta)}\cosh(2\pi\delta z)\cos(2\pi(x-ct))
\end{array}\label{solrot0}\ee with $c$ given by
(\ref{c}).

\section{Particle
trajectories}

Let $\left(x(t), z(t)\right)$ be the path of a particle in the
fluid domain, with location $\left(x(0), z(0)\right):=(x_0,z_0)$
at time $t=0$. Below
 small-amplitude capillary-gravity water
waves, taking into account (\ref{solrot0}), the motion of the
particles is described by the following differential system
  \be\left\{\begin{array}{ll}
 \f{dx}{dt}=u(x,z,t)=\f{2\pi\delta c}{\sinh (2\pi\delta)}\cosh (2\pi\delta
z)\cos(2\pi(x-ct))+c_0\\
 \f{dz}{dt}=v(x,z,t)=\f{2\pi c}{\sinh(2\pi\delta)}\sinh (2\pi\delta
z)\sin(2\pi(x-ct))
 \end{array}\right.\label{diff2}\ee
 The right-hand side of the differential system (\ref{diff2})
 is smooth and bounded; therefore, the unique solution of the Cauchy
 problem with initial data $(x_0,z_0)$ is defined globally in
 time.\\
 Notice that the constant $c_0$ is the average of the horizontal
fluid
 velocity over any horizontal
 segment of length 1; that is,
 \be
 c_0=\f 1 {1}\int_{x}^{x+1}u(s,z,t)ds,
 \ee
 representing therefore the strength of the underlying uniform
 current (see also \cite{CS}). Thus, $c_0=0$  will correspond to a region of still water with
 no underlying current,
 $c_0>0$ will characterize a favorable uniform current and  $c_0<0$
 will characterize an adverse uniform current.\\
To study the exact solution of  system (\ref{diff2}) it is
 more convenient to rewrite it in the following moving frame
 \be
 X=2\pi(x-ct),\quad  Z=2\pi\delta z \label{frame}
 \ee
This transformation yields \be\left\{\begin{array}{ll}
 \f{dX}{dt}=\f{4\pi^2\delta c}{\sinh(2\pi\delta)}\cosh(Z)\cos(X)+2\pi(c_0-c)\\
 \f{dZ}{dt}=\f{4\pi^2\delta c}{\sinh(2\pi\delta)} \sinh(Z)\sin(X)
 \end{array}\right.\label{diff3}\ee
\textbf{I)} $\mathbf{c_0=c}$\\
In this case,  differentiating with respect to $t$,  system
(\ref{diff3}) can be written into the following form:
\be\left\{\begin{array}{ll}
 \f{d^2 X}{dt^2}=-\f{8\pi^4\delta^2 c^2}{\sinh^2(2\pi\delta)}\sin(2X)\\
 \f{d^2Z}{dt^2}=\f{8\pi^4\delta^2 c^2}{\sinh^2(2\pi\delta)} \sinh(2Z)
 \end{array}\right.\label{30}\ee
 We denote by \be
A^2:= \f{8\pi^4\delta^2
 c^2}{\sinh^2(2\pi\delta)}
\label{a2}\ee We observe that $A^2$ as a function of $\delta$ is
decreasing and $ \lim_{\delta\rightarrow 0}A^2=2\pi^2c^2,\\
\lim_{\delta\rightarrow \infty}A^2=0$. Thus, \be 0\leq A^2\leq
2\pi^2c^2 \ee
System (\ref{30}) integrates at \be
\left\{\begin{array}{ll}
 \left(\f{dX}{dt}\right)^2=A^2
 \cos(2X)+c_1\\
 \left(\f{dZ}{dt}\right)^2=A^2
 \cosh(2Z)+c_2
 \end{array}\right.\label{31}
\ee $c_1$, $c_2$ being the integration constants. Because the
right-hand side of the first equation in (\ref{31}) has to be
bigger then zero, the constant $c_1$ has to satisfy the following
condition: \be c_1+A^2>0 \label{c1+a2}\ee For the first equation
in (\ref{31}) we use the substitution
 \be \tan(X)=y\, ,\,
 \cos(2X)=\f{1-y^2}{1+y^2}\,,\,\sin
(2X)=\f{2y}{1+y^2}\,, \, dX=\f{1}{1+y^2}dy\label{substitution}\ee
 In the new variable, the first equation in (\ref{31})
 takes the form
 \be \left(\f{dy}{dt}\right)^2=A^2(1-y^4)+c_1(1+y^2)^2\label{32}
 \ee
The solution of the equation (\ref{32}) involves an elliptic
integral of the first kind:
 \be \pm\int\f{dy}{\sqrt{(c_1-A^2)y^4+2c_1y^2+c_1+A^2}}=t\label{33}\ee The
elliptic integral of the first kind from (\ref{33}) may by reduced
to the Legendre normal form. In order to do this we consider the
substitution \be y^2=s \label{39} \ee Therefore, the left-hand
side in
 (\ref{33})
becomes \be \pm\int\f{dy}{\sqrt{(c_1-A^2)y^4+2c_1y^2+c_1+A^2}}=\pm
\int\f{ds}{2\sqrt{(c_1-A^2)s(s+1)\left(s+\f{c_1+A^2}{c_1-A^2}\right)}}
\label{40'}\ee Further, we introduce a new variable $\varphi$. The
definition of this variable depends on the sign of $c_1-A^2$.\\
If \be c_1-A^2>0 \ee then we introduce the variable $\varphi$ by
\be s=\tan^2\varphi\label{varphi}
 \ee
and we get \ba
&&(c_1-A^2)s(s+1)\left(s+\f{c_1+A^2}{c_1-A^2}\right)=
(c_1+A^2)\f{\tan^2\varphi}{\cos^4\varphi}\left[
1-k_1^2\sin^2\varphi \right] \nonumber\\
&&ds=2\f{\tan\varphi}{\cos^2\varphi}d\varphi\nonumber\ea where the
constant $0<k_1^2<1$ is given by \be
k_1^2=\f{2A^2}{c_1+A^2}\label{k1}\ee Therefore we obtain the
Legendre normal form of the integral in (\ref{33}); that is, \be
\pm
\f{1}{\sqrt{c_1+A^2}}\int\f{d\varphi}{\sqrt{1-k_1^2\sin^2\varphi
}}=t \label{40}\ee The inverse of the integral in (\ref{40}) is
the Jacobian elliptic function \textit{sine amplitude} (see, for
example, \cite{byrd}), an odd periodic function of order two, \be
\textrm{ sn }\left(\pm\sqrt{c_1+A^2}\,t;k_1\right):
=\sin\varphi\ee In view of the notations (\ref{39}),
(\ref{varphi}), we get that \be y(t)=\pm \f{\textrm{ sn
}\left(\sqrt{c_1+A^2}\,t;k_1\right)}{\textrm{ cn
}\left(\sqrt{c_1+A^2}\,t;k_1\right)}:=\pm
\textrm{sc}\left(\sqrt{c_1+A^2}\,t;k_1\right)\label{sol}\ee where
cn$\left(\sqrt{c_1+A^2}\,t;k_1\right):=\cos\varphi$ is the
Jacobian elliptic function \textit{cosine amplitude}, an even
periodic function of order two, and sc is Glaisher's notation for
the quotient sn/cn (see, for example, \cite{byrd}).

 If \be
c_1-A^2<0 \ee then, taking also into account the condition
(\ref{c1+a2}), we introduce the variable $\varphi$ by (see
\cite{smirnov} Ch. VI, \S 4, page 602) \be
s=\f{A^2+c_1}{A^2-c_1}\cos^2\varphi\label{varphi'} \ee In this
case  we get (see \cite{io3}) \be y(t)=\pm \sqrt{\f
{A^2+c_1}{A^2-c_1}}\,\textrm{cn
}\left(\sqrt{2A^2}\,t;k_2\right)\label{sol'}\ee where the constant
$0<k_2^2<1$ is given by \be k_2^2=\f{A^2+c_1}{2A^2}\ee

 For the second equation in (\ref{31})
 we use the substitution
 \be \tanh(Z)=w\, ,\quad
 \cosh(2Z)=\f{1+w^2}{1-w^2}\,, \quad
dZ=\f{1}{1-w^2}dw\label{substitution'}\ee
 In the new variable, the second equation in (\ref{31}) takes the form
 \be
\left(\f{dw}{dt}\right)^2=A^2(1-w^4)+c_2(1-w^2)^2\label{32'}
 \ee
The solution of the equation (\ref{32'}) involves an elliptic
integral of the first kind:
 \be \pm\int\f{dw}{\sqrt{(c_2-A^2)w^4-2c_2w^2+c_2+A^2}}=t\label{33'}\ee
  The
elliptic integral of the  first kind from (\ref{33'}) may by
reduced to the Legendre normal form. In order to do this we
consider the substitution \be w^2=r \label{39'} \ee The left-hand
side in
 (\ref{33'})
becomes \be
\pm\int\f{dw}{2\sqrt{(c_2-A^2)w^4-2c_2w^2+c_2+A^2}}=\pm
\int\f{dr}{2\sqrt{(c_2-A^2)r(r-1)(r-\f{c_2+A^2}{c_2-A^2})}}
\label{401}\ee As in the case of the  integral in (\ref{40'}), we
introduce a new variable $\phi$.
 The definition of  $\phi$ depends on the sign of $c_2-A^2$ and
$c_2+A^2$. There are three possibilities:\\
 $c_2-A^2>0$,\\
$c_2-A^2<0$ and $c_2+A^2>0$,\\
$c_2+A^2<0$. \\

If \be c_2-A^2>0 \ee then  we introduce the variable $\phi$ by
(see \cite{smirnov} Ch. VI, \S 4, page 602) \be
r=\sin^2\phi\label{varphi1}
 \ee
and we get \ba
&&(c_2-A^2)r(r-1)\left(r-\f{c_2+A^2}{c_2-A^2}\right)=
(c_2+A^2)\sin^2\phi\cos^2\phi\left(1-k^2_3\sin^2\phi
\right) \nonumber\\
&&dr=2\sin\phi\cos\phi d\phi\nonumber\ea where the constant
$0<k_3^2<1$ is given by \be k_3^2=\f{c_2-A^2}{c_2+A^2}\ee
Therefore we obtain the Legendre normal form of the integral in
(\ref{33'}); that is, \be \pm
\f{1}{\sqrt{c_2+A^2}}\int\f{d\phi}{\sqrt{1-k_3^2\sin^2\phi }}=t
\label{401'}\ee The inverse of the integral in (\ref{401'}) is the
Jacobian elliptic function sn \be \textrm{ sn
}\left(\pm\sqrt{c_2+A^2}\,t;k_3\right): =\sin\phi\ee In view of
the notations (\ref{39'}), (\ref{varphi1}), we get that \be
w(t)=\pm\textrm{ sn
}\left(\sqrt{c_2+A^2}\,t;k_3\right)\label{sol1}\ee If \be
c_2-A^2<0 \quad \textrm{ and }\quad c_2+A^2>0\ee then  we
introduce the variable $\phi$ by (see \cite{smirnov} Ch. VI, \S 4,
page 602) \be r=\cos^2\phi\label{varphi1'}
 \ee
and we get \ba
&&(c_2-A^2)r(r-1)\left(r-\f{c_2+A^2}{c_2-A^2}\right)=
2A^2\sin^2\phi\cos^2\phi\left(1-k^2_4\sin^2\phi
\right) \nonumber\\
&&dr=-2\sin\phi\cos\phi d\phi\nonumber\ea where the constant
$0<k_4^2<1$ is given by \be k_4^2=\f{A^2-c_2}{2A^2}\ee Therefore
we obtain the Legendre normal form of the integral in (\ref{33'});
that is, \be \pm
\f{1}{\sqrt{2}A}\int\f{d\phi}{\sqrt{1-k_4^2\sin^2\phi }}=t
\label{401''}\ee The inverse of the integral in (\ref{401''}) is
\be \textrm{ sn }\left(\pm\sqrt{2A^2}\,t;k_4\right): =\sin\phi\ee
In view of the notations (\ref{39'}), (\ref{varphi1'}), we get
that \be w(t)=\pm\textrm{ cn
}\left(\sqrt{2A^2}\,t;k_4\right)\label{sol1'}\ee

If \be c_2+A^2<0\ee then  we introduce the variable $\phi$ by (see
\cite{smirnov} Ch. VI, \S 4, page 602) \be r=
1+\f{2A^2}{c_2-A^2}\sin^2\phi \label{varphi1''}
 \ee
 In this case we get (see \cite{io3})
 \be
w(t)=\pm\sqrt{1-\f{2A^2}{A^2-c_2}\textrm{ sn
}^2\left(\sqrt{A^2-c_2}\,t;k_5\right)}\label{sol1''}\ee where the
constant $0<k_5^2<1$ is given by \be k_5^2=\f{2A^2}{A^2-c_2}\ee

 Thus, with (\ref{substitution}) and (\ref{substitution'})
 in view,
 the solution of  system (\ref{31}) has the following
expression  \be
\begin{array}{ll}
 X(t)=\textrm{arctan }[y(t)]\\
 Z(t)=\textrm{arctanh }[w(t)]=\f1{2}\ln\f{1+w(t)}{1-w(t)}
 \end{array}\label{34}
\ee with $y(t)$ given by (\ref{sol}) or (\ref{sol'})  and $w(t)$
given by (\ref{sol1}) or (\ref{sol1'}) or (\ref{sol1''}).
 From (\ref{frame}) and  (\ref{34}), the solutions of
system (\ref{diff2}) with the constant $c_0$ equal to  the speed
of propagation of the linear wave $c$ have the following
expressions: \be \left\{
\begin{array}{ll}
 x(t)=ct\pm\f1{2\pi}\textrm{arctan }\left[\textrm{sc}\left(\sqrt{c_1+A^2}\,t;
\f{2A^2}{c_1+A^2}\right) \right]\\
 z(t)=\pm\f 1{2\pi\delta}\textrm{arctanh }\left[\textrm{sn}\left(\sqrt{c_2+A^2}\,t;\f{c_2-A^2}{c_2+A^2}\right)\right]
 \end{array}\right.\label{solutie1}
\ee

 \be\left\{
\begin{array}{ll}
 x(t)=ct\pm\f1{2\pi}\textrm{arctan }\left[\textrm{sc}\left(\sqrt{c_1+A^2}\,t;
\f{2A^2}{c_1+A^2}\right)\right]\\
 z(t)=\pm\f 1{2\pi\delta}\textrm{arctanh }\left[\textrm{cn}\left(\sqrt{2A^2}\,t;\f{A^2-c_2}{2A^2}\right)\right]
 \end{array}\right.\label{solutie2}
\ee

\be\left\{
\begin{array}{ll}
 x(t)=ct\pm\f1{2\pi}\textrm{arctan }\left[\textrm{sc}\left(\sqrt{c_1+A^2}\,t;
\f{2A^2}{c_1+A^2}\right)\right]\\
 z(t)=\pm\f 1{2\pi\delta}\textrm{arctanh }\left[
 \sqrt{1-\f{2A^2}{A^2-c_2}\textrm{ sn}^2\left(\sqrt{A^2-c_2}\,t;\f{2A^2}{A^2-c_2}\right)}\right]
 \end{array}\right.\label{solutie3}
\ee

\be\left\{
\begin{array}{ll}
 x(t)=ct\pm\f1{2\pi}\textrm{arctan }\left[ \sqrt{\f
{A^2+c_1}{A^2-c_1}}\textrm{ cn}\left(\sqrt{2A^2}\,t;\f{c_1+A^2}{2A^2}\right)\right]\\
 z(t)=\pm\f 1{2\pi\delta}\textrm{arctanh }\left[\textrm{sn}\left(\sqrt{c_2+A^2}\,t;\f{c_2-A^2}{c_2+A^2}\right)\right]
 \end{array}\right.\label{solutie4}
\ee

\be\left\{
\begin{array}{ll}
 x(t)=ct\pm\f1{2\pi}\textrm{arctan }\left[ \sqrt{\f
{A^2+c_1}{A^2-c_1}}\textrm{ cn}\left(\sqrt{2A^2}\,t;\f{c_1+A^2}{2A^2}\right)\right]\\
 z(t)=\pm\f 1{2\pi\delta}\textrm{arctanh }\left[\textrm{cn}\left(\sqrt{2A^2}\,t;\f{A^2-c_2}{2A^2}\right)\right]
 \end{array}\right.\label{solutie5}
\ee

 \be\left\{
\begin{array}{ll}
 x(t)=ct\pm\f1{2\pi}\textrm{arctan }\left[ \sqrt{\f
{A^2+c_1}{A^2-c_1}}\textrm{ cn}\left(\sqrt{2}A\,t;\f{c_1+A^2}{2A^2}\right)\right]\\
 z(t)=\pm\f 1{2\pi\delta}\textrm{arctanh }\left[\sqrt{1-\f{2A^2}{A^2-c_2}\textrm{ sn}^2\left(\sqrt{A^2-c_2}\,t;\f{2A^2}{A^2-c_2}\right)}\right]
 \end{array}\right.\label{solutie6}
\ee Let us analyze  in more detail the solution (\ref{solutie1}).
Taking into account the expressions for the derivatives of sine
amplitude and cosine
 amplitude (see, for example, \cite{byrd}), that is,
 \ba&&\f {d}{dt}\textrm{ sn }(t;k)=\textrm{ cn }(t;k)
 \textrm{ dn }(t;k)\nonumber\\
 &&\f {d}{dt}\textrm{ cn }(t;k)=-\textrm{ sn }(t;k)\textrm{ dn }(t;k),\ea
where \be \textrm{ dn }(t;k):=\sqrt{1-k^2\textrm{ sn }^2(t;k)},
\ee
 we get
   the derivatives with
respect to $t$
  of $x(t)$ and
$z(t)$ from (\ref{solutie1}): \be\left\{
\begin{array}{ll}
x'(t)=c\pm\f{\sqrt{c_1+A^2}}{2\pi}\textrm{dn}\left(\sqrt{c_1+A^2}\,t;
\f{2A^2}{c_1+A^2}\right)\\
 z'(t)=\pm\f{\sqrt{c_2+A^2}}{2\pi\delta}\f{\textrm{dn}\left(\sqrt{c_2+A^2}\,t;
\f{c_2-A^2}{c_2+A^2}\right)}{\textrm{cn}\left(\sqrt{c_2+A^2}\,t;
\f{c_2-A^2}{c_2+A^2}\right)}
\end{array}\right.\label{solutie1'}\ee
For the alternative  with "+" in the expression (\ref{solutie1'})
of $x'(t)$, we obtain that $x'(t)>0$, for all $t$. The sign of the
derivative $z'(t)$ from (\ref{solutie1'}) depends on the sign of
the periodic Jacobian elliptic function cn. Thus,  we get, for
example, \be
 \begin{array}{ll}
 x'(t)>0,\,\, z'(t)<0 \quad \textrm{ for } \textrm{cn}\left(\sqrt{c_2+A^2}\,t;
\f{c_2-A^2}{c_2+A^2}\right)<0\\
 x'(t)>0,\,\, z'(t)>0 \quad \textrm{ for } \textrm{cn}\left(\sqrt{c_2+A^2}\,t;
\f{c_2-A^2}{c_2+A^2}\right)>0
 \end{array}\ee
 In this case, is the particle trajectory (\ref{solutie1}) an undulating curve to
 the right?\\
We observe that  for that $t$'s, denoted $\tilde{t}+K$, with $K$ a
period, for which the periodic Jacobian elliptic function sn
$\left(\sqrt{c_2+A^2}\,t; \f{c_2-A^2}{c_2+A^2}\right)=\pm 1$,
 we have
\be \lim_{t\rightarrow (\tilde{t}+K)}x(t)=\textrm{ finite
}:=\tilde{x}+K,\quad \lim_{t\rightarrow
(\tilde{t}+K)}z(t)=\pm\infty \label{limit'}\ee Therefore, at
$x=\tilde{x}+K$ the graph of the curve (\ref{solutie1}) will be
asymptotic. \\
Using Mathematica, for example, for $\delta=1$,  $c=10$, by
(\ref{a2}) we get $A^2=1.08704$, and  choosing  $c_1=7.91296>A^2$,
$c_2=2.91296>A^2$,
 the graph of the curve (\ref{solutie1})  with "+" in the expressions of $x(t)$ and $z(t)$,
 is drawn in   Figure 1.
For $\delta=\f {1}{2}$, $c=10$, from (\ref{a2}) we get
$A^2=146.07$, and  with  $c_1=177.93>A^2$, $c_2= 253.93>A^2$, the
graph of the curve (\ref{solutie1}) with "+" in the expressions of
$x(t)$ and $z(t)$,
 looks like in Figure 2.

\hspace{0cm}\scalebox{0.65}{\includegraphics{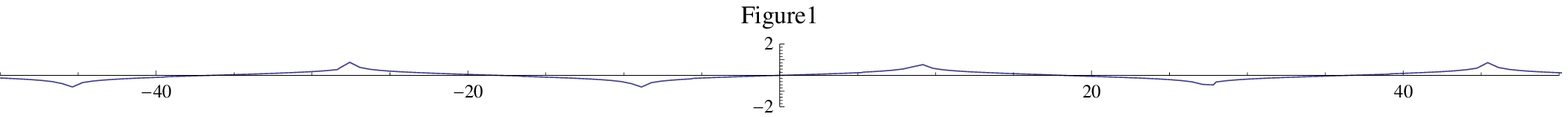}}

\hspace{0cm}\scalebox{0.65}{\includegraphics{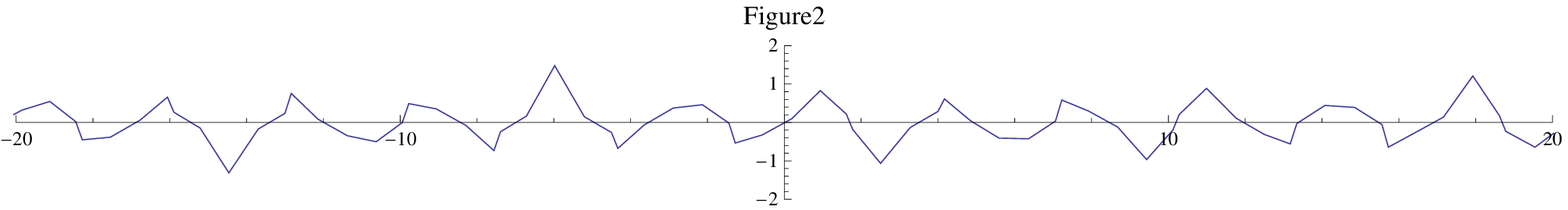}}

For the alternative with "-" in the expression (\ref{solutie1'}) of
$x'(t)$  we obtain that \\
\be
 \begin{array}{llll}
  \textrm{ if } c_1+A^2<4\pi^2c^2  \textrm{ then } x'(t)>0 \textrm{ for  all } t \\
 \textrm{ if } c_1-A^2>4\pi^2c^2 \textrm{ then } x'(t)<0 \textrm{ for  all } t\\
 \textrm{ if } c_1-A^2<4\pi^2c^2<c_1+A^2 \textrm{ then }\\
\hspace*{2.5cm}   x'(t)<0 \textrm{ for }   |\textrm{sn}\left(\sqrt{c_1+A^2}\,t;
\f{2A^2}{c_1+A^2}\right)|<\sqrt{\f{c_1+A^2-4\pi^2c^2}{2A^2}}\\
\hspace*{2.5cm}  x'(t)>0 \textrm{ for } |\textrm{sn}\left(\sqrt{c_1+A^2}\,t;
\f{2A^2}{c_1+A^2}\right)|>\sqrt{\f{c_1+A^2-4\pi^2c^2}{2A^2}}
\end{array}\ee
The sign of the derivative $z'(t)$ from (\ref{solutie1'}) depends
on the sign of the periodic Jacobian elliptic function cn; that
is, the sign of the derivative $z'(t)$ alternates successively,
$z'(t)<0$ and $z'(t)>0$.\\
 In this case, is the particle
trajectory (\ref{solutie1}) an undulating curve to the right, or
an undulating curve to the left, or a looping curve?  \\
Using Mathematica,  for $\delta=\f {1}{2}$, $c=10$, $A^2=146.07$,
$c_1=3822.93>A^2$ which satisfies $c_1-A^2<4\pi^2c^2<c_1+A^2$ and
$c_2= 2353.93>A^2$, the graph of the curve (\ref{solutie1}) with
"-" in the expression of $x(t)$ and "+" in the expression of
$z(t)$,
 is drawn in Figure 3.

\hspace{0cm}\scalebox{0.75}{\includegraphics{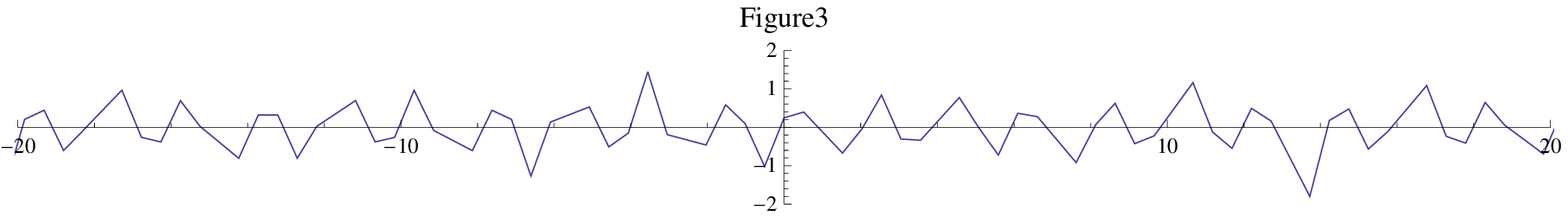}}

Using Mathematica, one can also draw the other solutions
(\ref{solutie2})-(\ref{solutie6}). For example, for
 $\delta=\f {1}{2}$, $c=10$, $A^2=146.07$,
$-A^2<c_1=46.07<A^2$ and $c_2=-253.93<-A^2$, the graph of the
curve (\ref{solutie6}) with "+" in the expressions of $x(t)$ and
$z(t)$, is drawn in the figure below.

\hspace{0cm}\scalebox{0.75}{\includegraphics{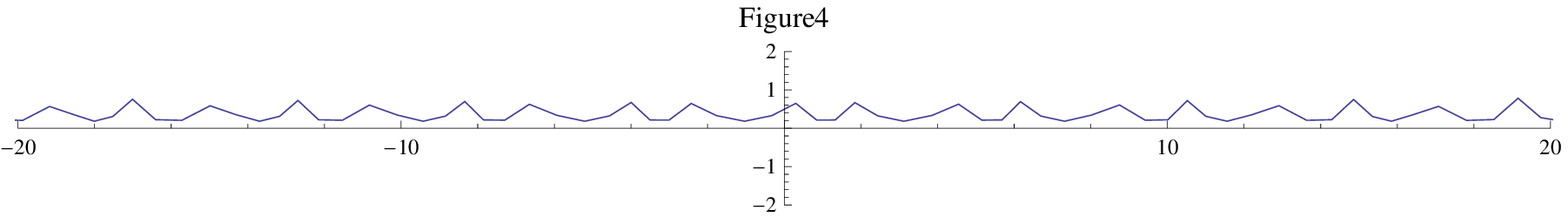}}

We remark that the curves obtained are not closed curves.

\textbf{II)} $\mathbf{c_0\neq c}$\\
Differentiating  system (\ref{diff3}) with respect to $t$, we get
\be
 \f{d^2 X}{dt^2}+b\tan(X)\f{dX}{dt}+A^2\sin(2X)-
 b^2\tan(X)=0\label{36}\ee
where $A^2$ is the constant from (\ref{a2}) and \be
b:=2\pi(c_0-c)\ee Using the substitution  (\ref{substitution}),
the equation
 (\ref{36})
takes the form \be \f{d^2
 y}{dt^2}-\f{2y}{1+y^2}\left(\f{dy}{dt}\right)^2+by\f{dy}{dt}+
 2A^2y-b^2y(1+y^2)=0
\label{37}\ee This differential equation can be written as an Abel
differential equation of  the second kind (see \cite{io3}). It is
solvable and its solution has the parametric form (for more
details, see \cite{io3}): \be
y(\tau)=\pm\sqrt{\f{\tau^2-2A^2}{\left(
C-b\ln|\tau+\sqrt{\tau^2-2A^2}|\right)^2}-1}, \label{ytau}\ee $C$
is a constant, and the relation between
 $t$ and $\tau$ is the following: \be t=\int
\f{1}{\sqrt{\tau^2-2A^2}\sqrt{\tau^2-2A^2-(C-b\ln|\tau+\sqrt{\tau^2-2A^2}|)^2}}\,d\tau
\label{t}\ee
 Thus, taking into account (\ref{substitution}), we obtain
\be X(t)=\textrm{arctan }[y(t)],\label{41}\ee with $y(\tau)$ given
by (\ref{ytau})
and $\tau$ given implicitly by (\ref{t}).\\
 In
order to determine $Z(t)$ from  system (\ref{diff3}), with
(\ref{41}) in view, we write the second equation of this system in
the form \be \f{dZ}{\sinh(Z)}=\f{4\pi^2\delta
c}{\sinh(2\pi\delta)}\sin(\textrm{arctan }[y(t)])\,dt
=\f{4\pi^2\delta
c}{\sinh(2\pi\delta)}\f{y(t)}{\sqrt{1+y^2(t)}}\,dt\ee
 If \be \int
\f{4\pi^2\delta
c}{\sinh(2\pi\delta)}\f{y(t)}{\sqrt{1+y^2(t)}}\,dt+\textrm{const}<
0 \label{int}\ee then we get (see \cite{io3}) \be
Z(t)=2\textrm{arctanh }\left[\exp\left(\int \f{4\pi^2\delta
c}{\sinh(2\pi\delta)}\f{y(t)}{\sqrt{1+y^2(t)}}\,dt+\textrm{const}\right)\right]
\label{42}\ee From (\ref{frame}), (\ref{41}) and (\ref{42}), the
solution of  system (\ref{diff2}) is written now as
\be\begin{array}{ll}
 x(t)=ct+\f1{2\pi}\textrm{arctan }\left[y(t)\right]\\
 z(t)=\f 1{\pi\delta}\textrm{arctanh }\left[\exp\left(\int
\f{4\pi^2\delta
c}{\sinh(2\pi\delta)}\f{y(t)}{\sqrt{1+y^2(t)}}\,dt+\textrm{const}\right)\right]
\end{array}\label{solutie''} \ee with $y(\tau)$ given by (\ref{ytau})
and $\tau$ given implicitly by (\ref{t}). Taking into
account(\ref{ytau}), (\ref{t}) and (\ref{a2}), we get \be \int
\f{4\pi^2\delta
c}{\sinh(2\pi\delta)}\f{y(t)}{\sqrt{1+y^2(t)}}\,dt=\f
1{2}\log\Big|\f{\tau(t)-\sqrt{2}A}{\tau(t)+\sqrt{2}A}\Big| \ee
Thus, the solution in (\ref{solutie''}) has, as function of the
parameter $\tau$, the following form: \be\begin{array}{lll}
 x(\tau)=c\int
\f{1}{\sqrt{\tau^2-2A^2}\sqrt{\tau^2-2A^2-(C-b\ln|\tau+\sqrt{\tau^2-2A^2}|)^2}}\,d\tau\\
\hspace{2cm}\pm\f1{2\pi}\textrm{arctan
}\left[\sqrt{\f{\tau^2-2A^2}{\left(
C-b\ln|\tau+\sqrt{\tau^2-2A^2}|\right)^2}-1}\right]\\
 z(\tau)=\pm\f {\textrm{const}}{\pi\delta}
 \textrm{arctanh }
 \left[\sqrt{\f{\tau-\sqrt{2}A}{\tau+\sqrt{2}A}}\right]
\end{array}\ee

 We also remark that the above curve is not a
closed curve.

\section{Acknowledgments}
I would like to thank Prof. A. Constantin for very helpful
comments and suggestions.

\medskip

\medskip

\end{document}